\documentclass[a4paper]{mem}
\usepackage{natbib}
\usepackage{graphicx}
\usepackage[a4paper]{hyperref}
\begin{document}

\newcommand{\etal}{{\rm et al.~}}
\newcommand{\be}{\begin{equation}}
\newcommand{\ee}{\end{equation}}
\def\bea{\begin{eqnarray}}
\def\eea{\end{eqnarray}}
\newcommand\eps{\varepsilon_{SN}}  
\newcommand{\cs}{c_{\star}}  
\newcommand{\tc}{{$t_{solve}$~}}  
\newcommand{\epg}{\varepsilon_{g}}  
\newcommand{\msu}{M_{\odot}}  
\newcommand{\Af}{{$A_{Fe}$}}
\newcommand{\La}{{$\Lambda$CDM}}
\newcommand{\gr}{\kern 2pt\hbox{}^\circ{\kern -2pt K}} %  ====> GRADI KELVIN
\newcommand{\oml}{\Omega_{\Lambda}}
\newcommand{\omm}{\Omega_{m}}
\newcommand{\brr}{\begin{array}}
\newcommand{\err}{\end{array}}
%\setcounter{table}{0}
%\setcounter{figure}{0}
% maggiore circa e minore circa%%%%%%%%%%%%%%%
\newcommand{\ltsima}{$\; \buildrel < \over \sim \;$}
\newcommand{\simlt}{\lower.5ex\hbox{\ltsima}}
\newcommand{\gtsima}{$\; \buildrel > \over \sim \;$}
\newcommand{\simgt}{\lower.5ex\hbox{\gtsima}}

   \title{Performance characteristics of a parallel treecode}

   \author{R. Valdarnini \inst{1} 
}

   \offprints{R. Valdarnini}

   \institute{SISSA via Beirut 2-4 34014 Trieste
 \email{valda@sissa.it} }

   \abstract{
I describe here the performances of a parallel treecode with individual 
particle timesteps. The code is  based on the Barnes-Hut algorithm and runs 
cosmological N-body simulations on parallel machines with a distributed memory 
architecture using the MPI message passing library. For a configuration with a 
constant number of particles per processor the scalability of the code has 
been tested up to $P=32$ processors. The average CPU time per processor 
necessary for solving the gravitational interactions is within  
$\sim 10 \%$ of 
that expected from the ideal scaling relation.
The load balancing efficiency is high ($\simgt90\%$) if the processor domains
are determined every large timestep according to a weighting scheme 
which takes into account the total particle computational load within the 
timestep.

   }
   \authorrunning{R. Valdarnini}
   \titlerunning{ Parallel treecode}
   \maketitle
%
%________________________________________________________________

\section{Introduction}

Numerical simulations play a fundamental role for improving the theoretical 
understanding of structure formation. 
This approach 
has received a large impulse from the huge growth of computer technology 
in the last two decades. Cosmological N-body simulations are now widely
used as a fundamental tool in modern cosmology for testing viable theories 
of  structure formation. 
 A popular approach for solving the gravitational forces of the system 
is the tree algorithm
\citep{ap85,he87}. The particle distribution of the system is arranged into 
a hierarchy of cubes and the force on an individual particle is computed 
by a summation over the multipole expansion of the cubes. 
An important point in favor of tree codes is that individual 
timesteps for all of the particles can be implemented easily, this allows a 
substantial speed-up  of the force evaluation for a clustered distribution. 

 An important task is the improvement of the dynamic range of the 
simulations. Large simulation volumes are required for statistical 
purposes, but at the same time modelling the formation and evolution 
of each individual galaxy in the simulated volume requires that a 
realistic simulation should be implemented with $10^8 \sim 10^9$ 
particles.
This computational task can be efficiently solved if the code is adapted 
to work on a parallel machine where many processors are linked together 
with a communication network.  
This has led a number of 
authors to parallelize treecodes \citep{sa91,wa94,du96,da97,li00,sp01,mi02}.
In this paper I present a parallel implementation of a multistep 
treecode based on the 
Barnes-Hut (1986, BH) algorithm. The code is cosmological and uses the MPI
message library.

\section{Parallelization of a treecode}
The BH algorithm works by subdividing a root box of size $L$, which contains 
all of the simulation particles, into 8 subvolumes of size $L/2$. This procedure
is then repeated for each of the subcubes and continues until the remaining 
cells or nodes
 are empty or have one particle. After the $k-th$ iteration the size 
of the subcubes is $l_k=L/2^k$. 
After the tree construction is complete
the multipole moments of the mass distribution inside the cells are 
computed starting from the smallest cells and  proceeding
 up to the root cell. 
The moments of the cells are typically approximated up to quadrupole order.
For each particle the acceleration is  
evaluated by summing the contribution of all of the cells  and particles
which are in an interaction list. The list is constructed starting
from the root cell and descending the tree down to a required 
level of accuracy. At each level a cell of the tree is accepted 
if it satisfies an accuracy criterion. 
If the cell fails this criterion then it is opened, the  
particles contained are added to the interaction list and the accuracy 
criterion is applied again for the remaining subcells.
The following acceptance criterion  has been used \citep{ba94,du96}
\be
d> l_k/\theta+\delta,
\label{opn}
\ee
where $d$ is the distance between the center of mass (c.o.m.) of the cell and 
the particle position, $\theta$ is an input parameter that controls the 
accuracy of the force evaluation, and $\delta$ is the distance between the 
cell c.o.m. and its geometrical center. 

\subsection{Domain decomposition}
 The spatial domains of the processors are determined according to the 
orthogonal recursion bisection (ORB, Salmon 1991). The computational volume is 
first cut along the x-axis at a position $x_c$ such that 
\be
\sum_{i<} w_i \simeq \sum_{i>} w_i,
\label{orb}
\ee
 where the summations are over all of the 
particles 
with $x_i <x_c$ or $x_i > x_c$ and $w_i\propto N_{OP}(i)$
 is a weight assigned to 
each particle proportional to the number of floating point operations
(i.e. the computational work) which are necessary to compute the particle 
force.

When the root $x_c$ has been determined the particles are then exchanged 
between the processors, until all of the particles with $x_i <x_c$ belong 
to the first $P/2$ processors and those with $x_i > x_c$ are in the second
 $P/2$ processors.
The whole procedure is repeated recursively, cycling through the cartesian
dimensions, until the total number of subdivisions of the computational 
volume is $log_2 P$ ( with this algorithm $P$ is constrained 
to be a power of two).
At the end of the domain decomposition the subvolumes will enclose a 
subset of particles with approximately an equal amount of computational 
work. The calculation of the forces is then approximately 
load-balanced among all of the processors.

\subsection{ Construction of the local essential tree}
A BH tree is constructed by each processor using the particles located in 
the processor subvolume. However, the local tree does not contain all of
the information needed to perform the force calculation for the 
processor particles.
For these particles a subset of cells must be imported from the trees of the
 other processors according to the opening angle criterion applied to the 
remote cells. 
Each processor then receives a set of partial trees which are merged with the 
local tree  to construct a local essential tree \citep{du96}. The new local
tree contains 
all of the information with which the forces of the local particles can be 
consistently calculated.

The communications between processors of nodes from different trees implies 
that in order to graft the imported cells onto the processor local trees 
it is 
necessary to adopt an efficient addressing scheme for the memory location of 
the nodes.
This is easily obtained if the construction of the local trees starts from a
root box of size $L$, common to all of the processors. 
The main advantage is that now the non-empty cells of
the local trees have the same position and size in all of the processors.
Each cell is then uniquely identified by a set of integers $\{j_1,j_2,...\}$,
with each integer ranging from $0$ to $7$ which identifies one of the $8$ 
subcells of the parent cell. These integers can be conveniently mapped 
onto a single integer word of maximum bit length $3k_{max}$, where 
$k_{max}$ is the maximum subdivision level of the tree. For a $64$ bit
key $k_{max} \le 21$.
This integer word represents the binary key of the cell.
When a cell of the tree is requested from a remote processor to construct its
local tree, the associated key is sent together with the mass, c.o.m. and 
multipole moments of the cell. The receiving processor then uses this key 
to quickly identify the cell location in the local tree and to add the new
cell to the local tree. A similar addressing scheme has been implemented, 
in their version of a parallel treecode, also 
by \citet{mi02}.
An efficient construction of the local essential tree is thus obtained as 
follows.

i) Once the ORB has been completed and each processor has received the 
particle subset with spatial coordinates within its spatial domains,
the local trees are constructed according to BH in each of the 
processors $P_k$, where $k$ is a processor index ranging from $0$ to $P-1$.
 
ii ) The communications between processors  can be 
significantly reduced if one adopts the following criterion to construct the 
partial trees that will be exchanged between processors.
After the local trees have been constructed, each processor applies 
the opening angle criterion between the nodes of its local tree and the 
closest point of the volume of another processor $P_k$.
The partial tree obtained contains by definition all the nodes of the local
processor necessary to evaluate the forces of the particles located in the
processor $P_k$. This procedure is performed at the same time by each
processor for all of the remaining $P-1$ processors. 
At the end, each processor has $P-1$ lists of nodes which are necessary for the 
construction of the local essential trees in the other processors.
The processor boundaries are determined during the ORB and are communicated 
between all of the processors after its completion.
Therefore the main advantage of this procedure is that all of the communications
between processors necessary for the construction of the local essential trees
are performed in a single all-to-all message passing routine. 
The drawback of this scheme is the memory overhead, because each
processor imports from another processor a list of nodes in excess of those 
effectively needed to perform the force calculation. 
As a rule of thumb it has been found that for $\theta=0.4$ a processor
with $N_p$ particles and $N_c$ cells imports $\sim N_p/8-N_p/4$ particles 
and $\sim N_c$ cells. The number of imported nodes is independent of 
the processor number. The value $\theta=0.4$ is a lower limit that guarantees
reasonable accuracy in the force evaluation in many simulations.
In the communication phase between processors 
mass and position are imported for each particle, and
 the mass, c.o.m., quadrupole moment and the binary key are imported for each
 cell. 

The memory required by a single processor to construct the local essential 
tree is then approximately a factor $\sim2$ larger than that used in the 
implementation of the local tree. This memory requirement can be efficiently 
managed with dynamic allocation, and is not significantly larger than that 
%a small of that globally necessary to the code.
required with other schemes used to construct the local essential tree
(e.g., Dubinski 1996).

\subsection{Force calculation }
After the construction of the local essential trees has been 
completed, each processor proceeds asynchronously to calculate 
up to the quadrupole order the forces of the active particles in its
computational volume. The code has incorporated periodic boundary 
conditions and comoving coordinates. Therefore the forces obtained
from the interaction lists of the local essential trees must be 
corrected to take into account the contribution of the images.
\citep{da97,sp01}.
These correction terms are calculated before the simulation
 using the Ewald method. 
The corrections are computed on a cubic mesh of
size $L$ with $50^3$ grid points and stored in a file.
During the force computation a linear interpolation is used to
calculate, from the stored values, the correction terms  corresponding to
the particle positions.

In a cosmological simulation the evaluation of the peculiar forces in 
the linear regime is subject to large relative errors. This  is
because for a nearly homogeneous distribution, the net force acting on 
a particle is the result of the cancellation of the large partial forces
determined from the whole mass distribution.
From a set of test simulations Dav\'{e} et al. (1997) found that in the 
linear regime, when 
$\theta=0.4$ and the cell moments are evaluated up to the quadrupole order, the 
relative errors in the forces are $\simlt 7\%$.
This problem is not present at late epochs, when the clustering of the 
simulation particles is highly evolved and even for $\theta \simeq1$
the relative errors in the forces are small ($\simlt 1\%$). 
This imposes in the simulation the necessity of varying $\theta$ according
to the clustering evolution, since the computational cost of evaluating
the forces with a small value of $\theta$ is wasted in the non-linear regime.
In this regime the forces can be evaluated with an accuracy as good as that 
obtained in the linear regime, though using an higher value of $\theta$.

After several tests it has been found that a good criterion to control the 
value of $\theta(t)$ is that at any given simulation time $t$ the 
energy conservation must be satisfied with a specified  level of 
accuracy. The Lyzer-Irvine equation is 
\be
a^4 T +aU -\int U da =C,
\label{enc}
\ee
where $a=a(t)$ is the expansion factor, $T$ is the kinetic energy of the 
system, $U$ is the potential energy and $C$ is a constant. The accuracy
of the integration can be measured by the quantity
$err(t)=|\Delta (C)/\Delta(aU)|$, where $\Delta f$ denotes the change of 
$f$ with
respect its initial value.
The time evolution of $err(t)$  has been analyzed 
 for different test simulations. 
The cosmological model considered is a flat CDM model, with a vacuum energy 
density $\oml=0.7$, matter density parameter $\omm=0.3$ and Hubble constant 
$h=0.7$ in units of $100 Km sec^{-1} Mpc^{-1}$. 
The power spectrum of the density fluctuations has been 
normalized so that the r.m.s. mass fluctuation in a sphere of 
radius $8 h^{-1} Mpc$ takes the value $\sigma_8=1$ 
at the present epoch, $a(t)=a_{fin}=11$. The simulations are run in a 
 $L=200 h^{-1}Mpc$ comoving box  with $N_p=84^3$ particles.
 For a simulation with $\theta=const=0.4$ 
 one has $err(t)\simlt 10^{-3}$ even in
 non-linear regimes, when $a(t)$ approaches its final value. 
The distribution of the relative root mean square errors in the 
force components of the particles  can be reproduced  if during the 
simulation $\theta=\theta(t)$ increases with time, provided that its value
never exceeds an upper limit
 implicitly defined for $\sigma_8 \geq 0.2$ 
by the constraint
\be
\Delta C /\Delta (aU) \leq 0.025 /[1+(0.4/\sigma_8)^{3}]^{1.7}.
\label{cvar}
\ee
An additional constraint sets an upper limit $\theta \leq 0.9$. 
This criterion can therefore be profitably used to constrain the
value of $\theta(t)$ according to the clustering evolution, and at 
the same time to maintain the relative errors in the forces below
a fixed threshold ($ \simlt 3\%$). This allows a substantial increase 
in the code performances. The computational cost of evaluating the forces 
depends on $\theta$ and for the considered runs 
 at $a(t)=11$  it
%clustered distribution at $a(t)=11$ it
is significantly reduced by a factor $10$ to $20$ when $\theta$ 
is increased from $0.4$ to $\sim0.9$.
\subsection{Multiple timesteps and particle update}
After the force calculation is complete, particle velocities and positions are 
updated in each processor. In the individual timestep scheme \citep{hk89} the 
particle timestep of particle $i$ is defined as 
$\Delta t_i = \Delta t_o/ 2^{n_i}$, where $n_i \geq0$ is an integer. 
The particle timesteps are determined according to several criteria.
The first is important at early  epochs and requires that  
\be 
\Delta t_i \leq \Delta t_{exp} =0.03 ~2/3~H(t),
\label{dt1}
\ee
where $H(t)$ is the Hubble parameter at the simulation time $t$.
The other two criteria are 
\begin{equation}
\Delta t_i \leq 0.3 ( \varepsilon_i~ a^3(t) /g_i)^{1/2}
\label{dt2}
\end{equation}
\begin{equation}
\Delta t_i \leq 0.3 ( \varepsilon_i/v_i)~,
\label{dt3}
\end{equation}
where $\varepsilon_i$ is the comoving gravitational softening parameter of 
the particle
 $i$, $g_i$ is the peculiar acceleration and $v_i$ its peculiar velocity.
These criteria are similar to those adopted by Dav\'e et al. (1997).
 At the beginning of the integration $t=t_{in}$, the forces are evaluated for 
all of the 
particles and their positions are advanced by half of the initial timestep,
which is common to all the particles.  
In this integration scheme the forces are evaluated at later times $t>t_{in}$ 
only for those particles for which is necessary to maintain the second-order
accuracy of the leapfrog integrator.
The particle positions are advanced using the smallest timestep 
$\Delta t_{min}$, as determined by the above constraints.
In the parallel implementation, each processor determines the individual
particle timesteps and the smallest timestep $\Delta t_{min}^{(pr)}$
of its particle subset, $\Delta t_{min}$ is then the smallest of these 
 $\Delta t_{min}^{(pr)}$
and is used by all the processors.

After that particle positions have been updated, it may happen that a
fraction of the particles assigned to a given processor have escaped
the processor subvolume.
At each timestep the particles that are not located within the original
processor boundaries are migrated between the processors.
 In principle the computational cost of locating the processor to which 
the escaped particle belongs
scales as the processor number $P$ ( Dav\'e et al. 1997, sect. 5.1). 
However, if during the ORB the processors are partitioned according to the 
procedure described in sect. 2.1, the final processor ordering 
makes it possible to reduce to $\sim \log_2 P$ the number of positional tests 
 of the particle. This is not a significant improvement
in pure gravity simulations, where the fraction of particles that 
leave a processor at each step is small ($ \sim 5\% $), but is important 
in a future implementation of the code that will incorporate 
smoothed particle hydrodynamic (Dav\'e et al. 1997, Springel et al. 2000). 
In such a scheme gas properties of a 
particle are estimated by averaging over a number of neighbors of the particle; 
 in a parallel implementation an efficient location of the particle 
neighbors located in the other processors is important in order to improve the
code performances.

   \begin{figure*}
   \centering
    \vspace{3cm}
   \includegraphics[width=10cm]{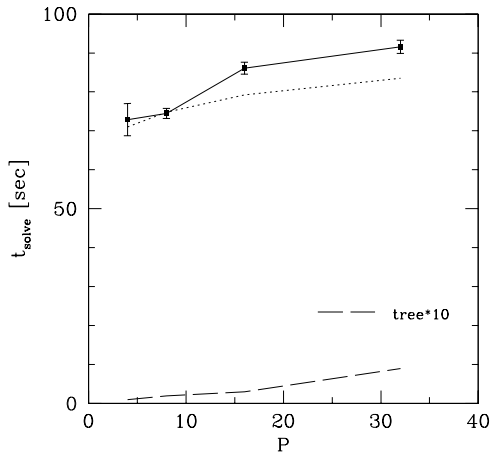}
    \vspace{-3cm}
\caption{
The averaged elapsed CPU time \tc spent in the force evaluation 
 as a function of the number of processors $P$. The value of the 
opening angle parameter is $\theta=0.4$, quadrupole moments are taken into
account.
The tests have been performed on an IBM SP3 machine, 
the error bars are the dispersions over the  $P$ processors.
 The total particle number of a test is $N_p=32^3P$. 
The dashed line is the CPU time
for constructing the tree scaled up by a factor 10. The initial particle
configuration is determined from a uniform distribution perturbed according to
a CDM power spectrum (see text). The dashed line is the expected \tc
from the ideal scaling relation $\propto log(N_p)$.
       }
         \label{tso}
   \end{figure*}

\section{Performances}
The treecode described here uses the BH tree algorithm to compute the 
gravitational forces. The implementation of this algorithm is identical 
to that of Dubinski (1996) and Dav\'e et al. (1997). 
The dependence of the errors in the force evaluations on a number of
input parameters has been discussed previously by these authors,
therefore an error analysis of the forces will not be presented here.
\subsection{Scalability}
The computational speed of the code is defined as the particle number divided
by the elapsed CPU wall-clock time \tc spent in the force computation of the 
particles.
 For a specified accuracy and particle distribution, 
the CPU time \tc 
of a parallel treecode with maximum theoretical efficiency 
is a fraction $1/P$ of that of the serial code.

The scalability of the parallel treecode has been tested by measuring \tc 
using a different number of processors $P$.
The initial positions of $32^3 P $ particles in a 
$L=11.11 h^{-1}Mpc$ comoving box have been perturbed according to a CDM model 
with $\omm=1$, $h=0.5$ and power spectrum normalization $\sigma_8=0.7$ at the
present epoch. 
The value of the  opening angle is $\theta=0.4$ and  forces have been 
computed at redshift $z=39$.
 The number of particles $N_p=32^3 P$ scales
linearly with $P$. This dependence of $N_p$ on $P$ has been chosen in
order to consistently compare the force solving CPU time \tc with
the one necessary for the construction of the local essential tree.
With the choice $N_p/P=const$ the CPU time \tc scales ideally 
as $t_{id}\propto \log N_p \propto \log P$. The results are shown in 
Fig. \ref{tso},
where \tc is plotted (continuous line) up to $P=32$.
For a configuration of $P$ processors \tc is defined as the average
of the values of the individual processors.
The tests have been performed on an IBM SP3 machine.
The dotted line shows the ideal scaling relation. In the large $P$ limit 
\tc is approximately  $10\%$ higher than the ideal scaling relation.
This is probably due to cache effects of the machine which arise 
 when $N_p \simgt 10^5$ during the tree descent necessary to calculate the 
forces.
An important result is the time $t_{tree}$ required to construct the local 
essential trees.
The dashed line of Fig. \ref{tso} shows that this time is always a small fraction 
($\simlt 5 \%$) of the time required to compute the gravity.
It is worth stressing that the communication part is efficiently handled by
the all-to-all routine and the corresponding time is a negligible fraction of
$t_{tree}$.
The computational speed is $ \sim 32^3/t_{solve} \sim 500P ~part/sec$.
This is valid for $\theta=0.4$. If $\theta$ is increased the particle 
interaction list will have a smaller number of terms and \tc will be
smaller. It has been found that $t_{solve}(\theta)$ is well  approximated by
 $t_{solve}(\theta \leq1)\propto10^{-5\theta/3}$.
If $\theta=1$ the CPU times of Fig. \ref{tso} will then be reduced by a factor 
$\sim 10$. A comparison of the code performances with those of other authors 
is difficult because of different algorithms, particle distributions and
machines. An educated guess that the code has performances fairly comparable
 with those of the Springel e al. code (2000) is given by Fig. 12 of their 
paper.
 In this figure the gravity speed as a function of the processor number is
shown for a cosmological hydrodynamic SPH simulation in a comoving box size
of $50 h^{-1} Mpc$ with $32^3$ dark matter particles and $32^3$ gas particles.
The cosmology is given by a \La~ model with $\omm=0.3$ and $h=0.7$. 
The simulations are evolved from an initial redshift $z_i=10$.
The plotted speeds have been measured on a CRAY T3E ($300MHz$ clock).
From Fig. 12 the computational speed of gravity for $P=32$ is 
$\sim 7500 part/sec$.
The cosmological model is not that adopted in the tests of Fig. \ref{tso}, but  
at early redshifts the computational cost of the gravity force calculation
is  not strongly dependent on the assumed model.
For $P=32$ and $\theta=0.4$ the results of Fig. \ref{tso} give a gravity speed 
of $\sim 12 \cdot 10^3 part/sec$ for a CDM model at $z_i=39$.
The measured speed must be reduced by  $\sim 20\%$ to take into account the 
higher clock rate of the IBM SP3 ($375 MHz$). The final value 
($\sim 9500 part/sec$) is similar to 
the one obtained by Springel et al. (2000).

   \begin{figure*}
   \centering
%    \vspace{3cm}
   \includegraphics[width=10cm]{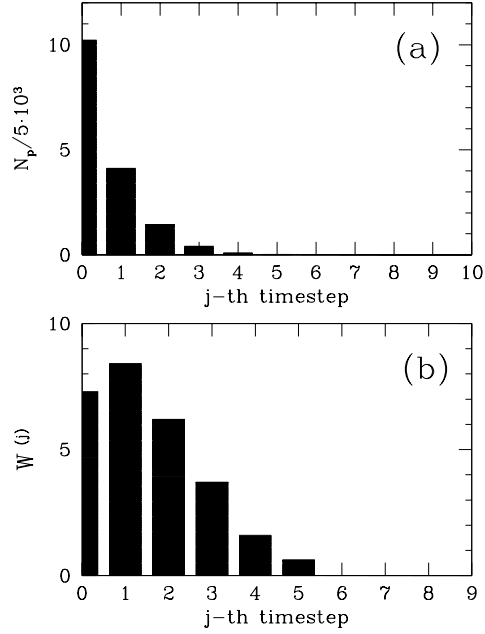}
\caption{ {\bf{(a):}} Number of particles with timesteps 
$\Delta t_j = \Delta t_0/2^j$
at the end of a macrostep $\Delta t_0$.
The simulation is that of sect. 3 with $N_p=2 \cdot 32^3$ particles, 
$\Delta t_0=t_{fin}/424$, $\Delta t_{min}=\Delta t_0/32$, and 
simulation time $t_n=n \Delta t_0=225 \Delta t_0$.
{\bf {(b):}} For the same particle distribution  
the corresponding computational loads 
$W^{(j)}=\sum _{i \in \Delta t_j} W_i$ are shown, the summation is over all the 
particles $i$ with timesteps $\Delta t_j$. 
  The individual particle weight used in the ORB  is
$W_{i} = \sum_k w_i^{(k)}$, here  $w_i^{(k)}$  
 is the
 single-step particle work at the $k-th$ step after $t_n$ and the summation
is over the steps between $t_n$ and $t_{n+1}$.
       }
         \label{whist}
   \end{figure*}

\subsection{Load balancing}
An important characteristic of a treecode is load balancing.
An ideal code should have the computational work divided evenly 
between the processors. This is not always possible and code performances
will be degraded when the load unbalancing between the processors becomes
large. At any point of the code where  
synchronous  communications are necessary there will be $P-1$ processors 
waiting for the 
most heavily loaded one to complete its computational work.
Load balancing can then be measured as
\be
L= \frac{1}{P} \sum_p 1 -(t_{max}-t_p)/t_{max},
\label{lbal}
\ee
where $t_p$ is the CPU time spent by the processor $p$ to complete a
procedure and $t_{max}$ is the maximum of the times $t_p$.
A treecode spends most of the CPU time in computing gravitational forces, and  
so it is essential to have good load balancing ($\simgt 90\%)$ with the 
gravity routine. 
This task is not obviously achieved with a multistep treecode.
The number of active particles for which it is necessary to compute the 
gravitational forces  at $t_n^{(k)}$ varies wildly with the timesteps. 
The current simulation time  
 $t_n^{(k)}$ 
is defined $k$ steps after $t_n= n \Delta t_0$ as 
$t_n^{(k)}=t_n +\sum_{j=1}^{j=k} 
\Delta t_j$, the summation is at $t > t_n$ over the past timesteps $\Delta t_j$.
At a certain step the ORB procedure described
in sect. 2.1 can be used to obtain load balancing, but at later steps the
unbalancing can be substantial. 
This problem has prompted some authors to consider more complicated 
approaches \citep{sp01,mi02}.
Here a simpler route is followed which starts from the observation
that in a multistep integration scheme, a better measure of the 
computational work done by each particle $i$ is given by 
$W_i =\sum _k w_i^{(k)}$, where $w^{(k)}_i \propto N_{OP}(i)$ is the 
number of floating point operations of particle $i$ necessary to calculate the 
gravitational forces of the particle at the simulation time $t_n^{(k)}$. 
If the particle $i$ is not active at $t_n^{(k)}$ , $w^k_i=0$.
The summation is over the steps 
between $t_n$ and $t_n + \Delta t_0$, the weights $W_i$ are now used  at
each large step $\Delta t_0$ to subdivide the computational volume
according to the ORB procedure.
Theoretically this weighting scheme does not guarantee a perfect 
load balance, nonetheless it has been found to yield satisfactory
results ($L \simgt 90 \%$) in many typical applications.
The reason lies in the shape of the distribution function $F(\Delta t_i)$ 
of the particle timesteps $\Delta t_i$, for a simulation with an evolved 
clustering state. 
The number of particles with 
timesteps in the interval $\Delta t_j, \Delta t_j+ \Delta t$ is
given by $n_j=F \Delta t$.   
The particle timesteps are determined according to the criteria 
defined in sect. 2.4; another parameter which determines the shape of
the distribution function is the minimum timestep $\Delta t_{min}$.

   \begin{figure*}
   \centering
%    \vspace{3cm}
   \includegraphics[width=10cm]{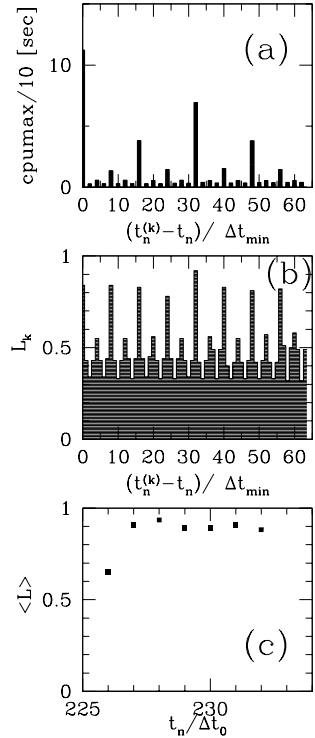}
\caption{ 
The load balancing scheme is tested for a parallel run with $P=4$ processors.
The simulation is that of Fig. \ref{whist}.  
{\bf{(a):}} 
The top panel shows, between $t_n$ and $t_{n+1}$, the maximum of the CPU times 
of the $P$  processors at the simulation time $t_n^{(k)}$, $n=225$.
The corresponding load balancing $L_k$ is plotted in the mid
panel {\bf {(b)}}.
The bottom panel shows the load balancing at the end of each macrostep
$\Delta t_0$. The goodness of the weighting scheme is shown by the first
point, where the ORB procedure has been performed setting $w_{i}=const$.
}
   \label{loadh}
   \end{figure*}

The optimal choice for $\Delta t_{min}$ requires that 
 the number of particles  
of the binned distribution $n_j$ in the last time bin should be  
 a small fraction of the total particle number
($N_{opt}\sim 10 \%N_p$). 
The distribution  $n_j$ of a test simulation is shown 
 as a function of the particle timesteps $\Delta t_j$ in Fig. \ref{whist}a 
at the end of a large timestep $\Delta t_0$, when
the particle positions are synchronized.
The simulation is that of sect. 3.1 with $N_p=2\cdot 32^3$ particles,
$\Delta t_o=t_{fin}/424$, $\Delta t_{min}=\Delta t_0/32$ and simulation 
time $t_n=n~ \Delta t_0=225~ \Delta t_0$.

The corresponding distribution of particle computational loads $W_i$ is
shown in panel (b). The plotted distribution is $W^{(j)}=\sum _{i \in
\Delta t_j} W_i$, the summation being over all of the particles $i$ with 
timesteps $\Delta t_j$. 
About $\sim 90\%$ of the particles are in the first three time bins, it can
be seen that for these bins the variations in the load distribution are 
within $\sim 20\%$. For example the number of particles $n_j$ 
with timestep $\Delta t_3=\Delta t_0/8$ is $\sim n_0 /10$.
The choice of a simple weighting scheme $w_i \propto N_{OP}(i)$ would 
have given a shape of the load distribution similar to that of $n_j$.
The reason for the shape of the load distribution of Fig. \ref{whist}b is that 
in a
multistep integration scheme the particle forces are calculated within 
a large timestep $\Delta t_0$ when their positions must be synchronized.
An optimal choice of the constraints on the particle timesteps 
yields a binned distribution $n_j$ with a hierarchy $n_{j+1}\sim n_j/2$.
A weighing scheme that sums the number of floating point operations
over a large timestep $\Delta t_0$ takes into account the fact there are
few particles with $\Delta t_j \ll \Delta t_0$ but that these 
particles have forces calculated a number of times $\propto \Delta t_j ^{-1}$.
This weighting scheme leads, at the end of a large timestep $\Delta t_0$, 
to particle loads with a distribution which can be considered for 
practical purposes roughly constant for a large fraction of the simulation 
particles ($\sim 90\%$). An ORB domain decomposition is then 
applied every large timestep $\Delta t_0$  according to the 
calculated weights of the particles.
The subdivision of the computational load that follows from this ORB
among the processors it is still unbalanced, but within a large timestep 
$\Delta t_0$ the unbalancing is higher when the computational work is
minimal. 
This is clearly illustrated in the example of Fig. \ref{loadh}.
The simulation of Fig. \ref{whist} has been used at the same 
simulation time, with $N_p=2\cdot 32^3 $ particles divided 
among $P=4$ processors. The processor domains have been
found as previously discussed. Panel (a) shows the elapsed CPU wall-clock
time spent by the parallel code to compute the gravitational 
forces.  The CPU time is plotted 
 between $t_n$ and $t_{n+1}$ versus the simulation time 
$t_n^{(k)}$ and is the maximum of  the single processor values.
There is a large burst of CPU work when the particles synchronized 
at $t_n^{(k)}$  are those with timesteps $\Delta t_0$, $\Delta t_0/2$ and
$\Delta t_0/4$.
The instantaneous load balancing $L_{(k)}$ is calculated using Eq. \ref{lbal} 
between $t_n^{(k)}$  and $t_n^{(k+1)}$. It can be seen that $L_{(k)}$
  drops to very inefficient values ($\simlt 0.3$) when 
 $t_n^{(k)}$  corresponds to a small number of active particles and it
reaches a high efficiency ($\simgt 0.9$) with the highest CPU times.
The overall load balancing is measured by applying Eq. \ref{lbal} 
between every  ORB domain decomposition: panel (c) shows $<L>$ versus 
the simulation time $t_n$ for ten large timesteps $\Delta t_0$.
To show how well the method is working the ORB procedure has been performed 
setting for the first step $w_i=const$, 
 yielding $<L>\sim 0.5$~.
This proves that the load balancing performances are sensitive to the 
chosen weighting scheme and that the procedure previously described is
optimal to achieve a good load balance for the parallel tree code 
described here. 

Preliminary results show that this weighting scheme for the ORB domain 
decomposition can still be successfully used 
 to obtain an efficient load balancing ($\simgt 90\%$) when  the number
of processors is higher than that of the tests performed here ($P=32$).
This confirms that the load balancing efficiency of the
adopted weighting scheme is robust for a variety of 
clustering states in cosmological simulations.

\bibliographystyle{aa}

\end{document}